\documentclass[12pt]{article}
\usepackage{amsfonts}
\usepackage{graphicx}
\usepackage{amsmath}
\usepackage{float}
\usepackage{amssymb}
\providecommand{\U}[1]{\protect\rule{.1in}{.1in}}
\textwidth=6.5in \hoffset=-.75in \voffset=-1in \numberwithin{equation}{section}
\numberwithin{figure}{section}

\begin{document}
\begin{titlepage}
\bigskip \begin{flushright}
\end{flushright}
\vspace{1cm}
\begin{center}
{\Large \bf {M-Branes on $k$-center Instantons}}\\
\vskip 1cm
\end{center}
\vspace{1cm}
\begin{center}
A. M.
Ghezelbash{ \footnote{ E-Mail: masoud.ghezelbash@usask.ca}}, R. Oraji
{ \footnote{ E-Mail: rao519@mail.usask.ca}}
\\
Department of Physics and Engineering Physics, \\ University of Saskatchewan,
Saskatoon, Saskatchewan S7N 5E2, Canada\\
\vspace{1cm}
\end{center}
\begin{abstract}

We present analytic solutions for membrane metric function 
based on transverse $k$-center instanton geometries.
The membrane metric functions depend on more than two transverse coordinates and  the solutions provide realizations of fully localized type IIA D2/D6 and NS5/D6 brane intersections. 
All solutions have partial preserved supersymmetries. 

\end{abstract}
\bigskip
\end{titlepage}\onecolumn

\bigskip

\section{Introduction}

Fundamental M-theory in the low-energy limit is generally believed to be
effectively described by $D=11$ supergravity \cite{gr1,gr2,gr3}. This
suggests that brane solutions in the latter theory furnish classical soliton
states of M-theory, motivating considerable interest in this subject. There
is particular interest in finding $D=11$ M-brane solutions that reduce to
supersymmetric $p$-brane solutions (that saturate the
Bogomol'nyi-Prasad-Sommerfield (BPS) bound) upon reduction to 10 dimensions.
Some supersymmetric BPS solutions of two or three orthogonally intersecting
2-branes and 5-branes in $D=11$\ supergravity were obtained some years ago 
\cite{Tsey}, and more such solutions have since been found \cite{oth}.

Recently interesting new supergravity solutions for localized D2/D6, D2/D4,
NS5/D6 and NS5/D5 intersecting brane systems were obtained \cite%
{CGMM2,ATM2,GHEbianchiMbranes,GH2resolvedconifolds}. By lifting a D6
(D5 or D4)-brane to four-dimensional self-dual geometries embedded in
M-theory, these solutions were constructed by placing M2- and M5-branes in
different self-dual geometries. A special feature of this construction\ is
that the solution is not restricted to be in the near core region of the D6
(or D5) brane, a feature quite distinct from the previously known solutions 
\cite{IT}. For all of the different BPS solutions, 1/4 of the
supersymmetry is preserved as a result of the self-duality of the transverse
metric. Moreover, in \cite{Ali}, partially localized D-brane systems
involving D3, D4 and D5 branes were constructed. By assuming a simple ansatz
for the eleven dimensional metric, the problem reduces to a partial
differential equation that is separable and admits proper boundary
conditions.

Motivated by this work, the aim of this paper is to construct the fully
localized supergravity solutions of D2 (and NS5) intersecting D6 branes
without restricting to the near core region of the D6 by reduction of ALE
geometries lifted to M-theory. 

In ref \cite{Rahim}, the authors obtained several different supersymmetric BPS solutions of interest, based on transverse embedded 2-center Gibbons-Hawking space.
All the solutions preserve eight supersymmetries and the metric functions depend on more than two transverse coordinates. The main motivation in this paper is extension of the results in \cite{Rahim},
to embed multi-center (and in particular three-center) 
Gibbons-Hawking space in M-theory. 

The outline of paper is as follows. In section \ref{sec:Msolutions}, we
discuss briefly 
the ALE geometries and present the eleven dimensional supergravity equations for M2-brane with an embedded transverse $k$-center instanton.

In section \ref{sec:Nnutcharge}, we present the solutions to membrane equations of motion for a transverse embedded $k$-center Gibbons-Hawking space where $r$ is greater than a multiple of $a$.
 
In sections \ref{sec:solutionsMiddle}, we present membrane solutions 
for an embedded $3$-center instanton and we find solutions in region $r>a$.
 
In section \ref{sec:M2mix}, we then discuss embedding products of
Gibbons-Hawking instantons in M2-brane solutions as well as $M_5$ brane solutions with one embedded Gibbons-Hawking instanton. We show all of the solutions presented in chapters 3, 4 and 5 preserve
some of the supersymmetry. 

In section \ref{sec:dec}, we consider the decoupling limit of our
solutions and find evidence that in the limit of vanishing string coupling,
the theory on the world-volume of the NS5-branes is a new little string
theory. Moreover, we apply T-duality transformations on type IIA solutions
and find type IIB NS5/D5 intersecting brane solutions and discuss the
decoupling limit of the solutions. We wrap up then by some concluding
remarks and future possible research directions.

\section{M-brane Solutions On $k$-center Instantons}
\label{sec:Msolutions}

We consider an M2-brane, given by the metric 
\begin{equation}
ds_{11}^{2}=H(y,r,\theta)^{-2/3}\left( -dt^{2}+dx_{1}^{2}+dx_{2}^{2}\right)
+H(y,r,\theta)^{1/3}\left( d\mathfrak{s}_{4}^{2}(y)+ds_{4}^{2}(r,\theta
)\right)  \label{ds11genM2}
\end{equation}
and four-form field strength 
\begin{align}
F_{tx_{1}x_{2}y} & =-\frac{1}{2H^{2}}\frac{\partial H}{\partial y}
\label{Fy} \\
F_{tx_{1}x_{2}r} & =-\frac{1}{2H^{2}}\frac{\partial H}{\partial r}
\label{Fr} \\
F_{tx_{1}x_{2}\theta} & =-\frac{1}{2H^{2}}\frac{\partial H}{\partial\theta}.
\label{Ft}
\end{align}

For an M5-brane, the metric reads as
\begin{equation}
ds^{2}=H(y,r,\theta)^{-1/3}\left( -dt^{2}+dx_{1}^{2}+\ldots+dx_{5}
^{2}\right) +H(y,r)^{2/3}\left( dy^{2}+ds_{4}^{2}(r,\theta)\right) ~~~
\label{ds11general} 
\end{equation}
and four-form field strength is 
\begin{equation}
F_{m_{1}\ldots m_{4}} =\frac{\alpha}{2}\epsilon_{m_{1}\ldots m_{5}
}\partial^{m_{5}}H,\label{Fgeneral}
\end{equation}

where $d\mathfrak{s}_{4}^{2}(y)$ and $ds_{4}^{2}(r,\theta)$ are two
four-dimensional (Euclideanized) metrics, depending on the non-compact
coordinates $y$ and $r$, respectively and the quantity $\alpha=\pm 1,$\ which
corresponds to an M5-brane and an anti-M5-brane respectively. The general
solution, where the transverse coordinates are given by a flat metric,
admits a solution with 16 Killing spinors \cite{smith}.
As it is well known, the 
metric of $k$-center $A$ series instantons could be
written in closed form,  
given by: 
\begin{equation}
ds^{2}=V^{-1}(dt+\vec A \cdot d \vec x)^{2}+V \gamma_{ij} dx^{i} \cdot d
x^{j}  \label{Aseries}
\end{equation}
where $V$, $A_{i}$ and $\gamma_{ij}$ are independent of $t$ and $\nabla
V=\pm\nabla\times\vec A $; hence $\nabla^{2} V=0$. The most general solution
for $V$ is then $V=\sum_{i=1}^{k} \frac{m}{\mid\vec x- \vec x _{i} \mid}$.
The metric (\ref{Aseries}) describes the Gibbons-Hawking multi-center
instantons. The $k=0$ corresponds to flat space and $k=1$ corresponds to
Eguchi-Hanson metric. The different M2 and M5 brane solutions with one (or two) transverse $k=2$ Gibbons-Hawking space have been constructed and studied extensively in \cite{Rahim}. In particular, the authors explicitly found exact supergravity solutions for fully localized D2/D6 and NS5/D6 brane
intersections without restricting to the near core region of the D6 branes.
The metric functions of all the solutions
depend on three (or four) transverse
coordinates. The common feature of all of these solutions is that the
brane function is a convolution of a decaying function with a damped
oscillating one. The metric functions vanish far from the M2 and M5 branes
and diverge near the brane cores.

In this paper we consider the extension of metrics (\ref{Aseries}) by considering 
\begin{equation}
V=\epsilon+\sum_{i=1}^{k} \frac{m_{i}}{\mid\vec x- \vec x _{i}
\mid}.  \label{Vepsilon}
\end{equation}
especially with $k=3$. The hyper-Kahler metrics (\ref{Aseries}) with $V$ pose a
translational self-dual (or anti-self-dual) Killing vector $K_{\mu}$, that
means 
\begin{equation}
\nabla_{\mu}K_{\nu}=\pm\frac{1}{2}\sqrt{det \, g}\epsilon_{\mu\nu}
^{\rho\lambda}\nabla_{\rho}K_{\lambda}.  \label{transkill}
\end{equation}
This (anti-) self-duality condition (\ref{transkill}) implies the
three-dimensional Laplace equation for $V$ with solutions (\ref%
{Vepsilon}). For $\epsilon\neq 0$ in (\ref{Vepsilon}), the metrics (\ref%
{Aseries}) describe the asymptotically locally flat (ALF) multi Taub-NUT
spaces. The removal of nut singularities implies $m_{i}=m$ and $t$ a
periodic coordinate of period $\frac{8\pi m}{k}$. 
We consider the Gibbons-Hawking space with $k=3$ and metric
function $V$ with $\epsilon \neq 0$, as a part of transverse
space to M2 and M5-branes. The four-dimensional Gibbons-Hawking metric with $k=N_1+N_2+1$ is 
\begin{equation}
ds_{GH}^{2}={V(r,\theta )}\{dr^{2}+r^{2}(d\theta ^{2}+\sin
^{2}\theta d\phi ^{2})\}+\frac{(d\psi +\omega (r,\theta )d\phi )^{2}}{%
V(r,\theta )}
\label{dsGH}
\end{equation}%
where 
\begin{align}
\omega (r,\theta )& =\sum_{m=-N_2}^{m=N_1}\frac{n(a+r\cos \theta )}{\sqrt{%
r^{2}+(ma)^{2}+2mar\cos \theta }}  \label{om} \\
V(r,\theta)&=\epsilon+\frac{n}{r}+\sum_{k=1}^{N_1}{\frac{n}{\sqrt{r^2+(ka)^2+2kar \cos \theta}}}+\sum_{k=1}^{N_2}{\frac{n}{\sqrt{r^2+(ka)^2-2kar \cos \theta}}}. 
\label{Vep}
\end{align}%
For later convenience, we define $\mathcal{N}=\max (N_1,N_2)$. The eleven dimensional M2-brane with an embedded transverse Gibbons-Hawking
space is given by the following metric 
\begin{equation}
ds_{11}^{2}=H(y,r,\theta )^{-2/3}\left( -dt^{2}+dx_{1}^{2}+dx_{2}^{2}\right)
+H(y,r,\theta )^{1/3}\left( dy^{2}+y^{2}d\Omega _{3}^{2}+ds_{GH}^{2}\right)
\label{ds11m2}
\end{equation}%
and non-vanishing four-form field components are given by eqs. (\ref{Fy}), (%
\ref{Fr}) and (\ref{Ft}). The metric (\ref{ds11m2}) is a solution to the
eleven dimensional supergravity equations provided $H\left( y,r,\theta
\right) $ is a solution to the differential equation 
\begin{align}
& 2ry\sin \theta \frac{\partial H}{\partial r}+y\cos \theta \frac{\partial H%
}{\partial \theta }+r^{2}y\sin \theta \frac{\partial ^{2}H}{\partial r^{2}}%
+y\sin \theta \frac{\partial ^{2}H}{\partial \theta ^{2}}+  \notag \\
& +(r^{2}y\sin \theta \frac{\partial ^{2}H}{\partial y^{2}}+3r^{2}\sin
\theta \frac{\partial H}{\partial \theta })V(r,\theta )=0.  \label{LapH}
\end{align}%
We notice that solutions to the harmonic equation (\ref{LapH}) determine the
M2-brane metric function everywhere except at the location of the brane
source. To maximize the symmetry of the problem, hence simplify the
analysis, we consider the M2-brane source is placed at the point $y=r=0$.
Separating the coordinates by taking 
\begin{equation}
H(y,r,\theta )=1+Q_{M2}Y(y)R(r,\theta )  \label{Hyrsep}
\end{equation}%
where $Q_{M2}$ is the charge on the M2-brane, the equation (\ref{LapH}) reduces to two separated differential
equations for $Y(y)$ and $R(r,\theta )$. The solution of the differential
equation for $Y(y)$ is 
\begin{equation}
Y(y)\sim \frac{J_{1}(cy)}{y}  \label{Y1}
\end{equation}%
which has a damped oscillating behavior at infinity. The differential
equation for $R(r,\theta )$ is 
\begin{equation}
2r\frac{\partial R(r,\theta )}{\partial r}+r^{2}\frac{\partial
^{2}R(r,\theta )}{\partial r^{2}}+\frac{\cos \theta }{\sin \theta }\frac{%
\partial R(r,\theta )}{\partial \theta }+\frac{\partial ^{2}R(r,\theta )}{%
\partial ^{2}\theta }=c^{2}r^{2}V(r,\theta )R(r,\theta )  \label{a0}
\end{equation}%
where $c$ is the separation constant.\\

\section{Supergravity Solutions for  $M_2$-Brane with Embedded $k$-center Instantons where $r>\mathcal{N}a$}
\begin{figure}[!ht]
\centering           
\begin{minipage}[c]{.6\textwidth}
        \centering
        \includegraphics[height=9.5cm]{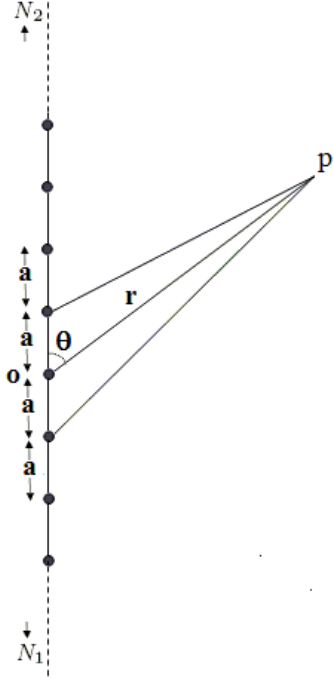}
    \end{minipage}
\caption{The geometry of charges in $k=N_1+N_2+1$-center instanton.}
\label{NNNcharges}
\end{figure}

{\label{sec:Nnutcharge} }
We try to find solutions to (\ref{a0}) in the presence of $k= N_1 + N_2 +1$ charges (Figure \ref{NNNcharges}) where the functional form of $V(r,\theta)$ is given by (\ref{Vep}).   

In general it is unlikely to find exact analtyic solutions to (\ref{a0}), hence we need to make some approximations. In this section and appendix A, we find the solutions of (\ref{a0}) in region $r>\mathcal{N}a$ and region $r<a$, respectively. \\

In region $r>\mathcal{N}a$, the metric function (\ref{Vep}) reduces to

\begin{equation}
V(r,\theta)\approx \epsilon+\frac{n(1+N_1+N_2)}{r}+ \left[ {\frac{N_2(N_2+1)-N_1(N_1+1)}{2}} \right ] \frac{an \cos \theta}{r^2}
\label{VNChargeA}
\end{equation}
where we keep the terms up to the second-order in $1/r$.

\begin{figure}[!ht]
\centering           
\begin{minipage}[c]{.6\textwidth}
        \centering
        \includegraphics[height=6.5cm]{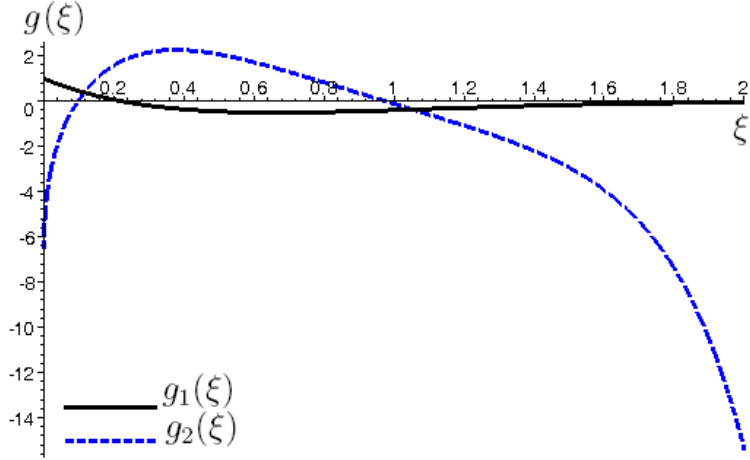}
    \end{minipage}
\caption{The first and second lines of solution (\ref{gg}) represented by $g_1(\xi)$ and $g_2(\xi)$, respectively. }
\label{NNNchargesF2}
\end{figure}

The separated differential equations after applying (\ref{VNChargeA}) are 
\begin{equation}
r^{2}\frac{d^{2}f(r)}{dr^{2}}+2r\frac{df(r)}{dr}-c^{2}(\epsilon
r^{2}+n(N_1+N_2+1)r+M^2)f(r)=0 
\label{NNa01}
\end{equation}
\begin{equation}
\frac{d^{2}g(\theta )}{d\theta ^{2}}+\frac{\cos \theta }{\sin \theta }\frac{%
dg(\theta )}{d\theta }+c^{2}(M^2+\tilde{m}\cos \theta )g(\theta )=0  
\label{NNa02}
\end{equation}
where\newline
\begin{equation}
\tilde{m}=\frac{(N_1 (N_1 +1)-N_2 (N_2 +1)}{2}na
\end{equation}
and the constants $c$ and $M$ are considered as real positive numbers.
\newline
The solution to equation (\ref{NNa01}) is given by
\begin{equation}
f(r)\sim\frac{1}{r}\mathcal{W}_{W}(-\frac{cn(N_1+N_2+1)}{2 \sqrt{\epsilon}}%
,\frac{\sqrt{1+4M^2c^{2}}}{2},2c{\sqrt{\epsilon}}r)
\label{NFa0}
\end{equation}
where $\mathcal{W}_{W}$ is a Whittaker function and the solution to equation (\ref{NNa02}) is given by
\begin{equation}
\begin{split}
g(\xi)=&C_{c,M} \mathcal{H}_{C}(0,0,0,2\tilde{m}c^{2},-(M^2+\tilde{m})c^{2},\frac{\xi}{2})+\\
&C'_{c,M}\mathcal{H}_{C}(0,0,0,2\tilde{m}c^{2},-(M^2+\tilde{m})c^{2},\frac{\xi}{2}) \int{\frac{d\xi}{\xi(\xi-2) {\mathcal{H}_{C}(0,0,0,2\tilde{m}c^{2},-(M^2+\tilde{m})c^{2},\frac{\xi}{2})}^2 }} 
\end{split}
\label{gg}
\end{equation}
where $\mathcal{H}_{C}$ is the Heun-C function (see appendix B), $\xi=1-\cos \theta$ and $C_{c,M},C'_{c,M}$ are constants. Figure (\ref{NNNchargesF2}) shows the behaviour of the first and second lines of (\ref{gg}) where the constants are set to $a=1$, $n=1$, $\tilde{m}=12 \ (N_1=5$ and $N_2=2)$, $M=1$, and $c=1$. As it's shown in appendix C, the second line of (\ref{gg}) has a logarithmic divergence at $\xi=1$.  
 \\
Knowing the general solution to (\ref{a0}), given by $R(r,\theta )=f(r){g(\xi)}$; we can write the membrane metric function as
\begin{equation}
H(y,r,\theta )=1+Q_{M2}\int_0^\infty dc \int _0 ^\infty dM Y(y)f(r){g(\xi)} 
  \label{NNa2}
\end{equation}
in region $r>\mathcal{N}a$.
As we notice, the solution (\ref{NNa2}) depends on two measure functions $C_{c,M}$ and $C'_{c,M}$. Each of these functions has dimension of inverse length to four. So, the measure functions should be considered as series expansions of the form $c^{\alpha+4}M^{\alpha}$ where $\alpha \in \mathbb {Z}_+$. 

In appendix A, the solutions to equation (\ref{a0}) are presented in other region of interest where $r<a$.
We are not able to find the analytic solutions in region $a<r<\mathcal{N}a$ for embedded $k$-center instantons where $k>3$. For $k=2$, the analytic solutions are already presented in \cite{Rahim} where $r$ takes any value $r\geq 0$. In next section, we consider the case of embedded $k=3$ center embedded Gibbons-Hawking space and we find the solutions on region $r >a$. 

\section{Supergravity Solutions for $M_{2}$-Brane with Embedded $3$-center Instantons where $r>a$}
{\label{sec:solutionsMiddle} }

\begin{figure}[!ht]
\centering           
\begin{minipage}[c]{.6\textwidth}
        \centering
        \includegraphics[height=6.5cm]{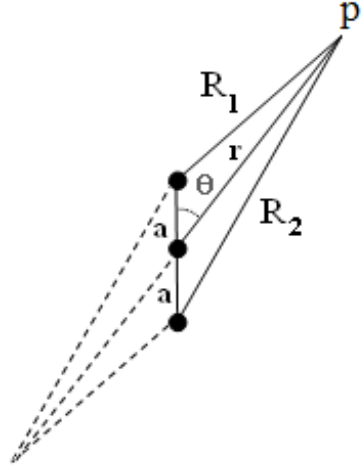}
    \end{minipage}
\caption{The relation between $\mu$, $\lambda$ and $r$.}
\label{p3mul}
\end{figure}

To find the solutions to (\ref{a0}) over region $r>a$, we define a  pair of new independent coordinates $\mu ,\lambda$ given by 
\begin{eqnarray}
\mu  &=&\frac{R_{2}+R_{1}}{2}=\frac{\sqrt{r^{2}+a^{2}+2ar\cos \theta }+\sqrt{%
r^{2}+a^{2}-2ar\cos \theta }}{2}  \label{vchagne} \\
\lambda  &=&\frac{R_{2}-R_{1}}{2}=\frac{\sqrt{r^{2}+a^{2}+2ar\cos \theta }-%
\sqrt{r^{2}+a^{2}-2ar\cos \theta }}{2}.  \label{vchange1}
\end{eqnarray}%
 A geometrical interpretation of $\mu$ and $\lambda$ can be obtained using Figure (\ref{p3mul}). According to Figure (\ref{p3mul}) we can easily show that $|R_2-R_1| < 2r <(R_1+R_2)$ and $|R_2-R_1| < 2a <(R_1+R_2)$ or in other words $\lambda < r < \mu$ and $\lambda < a < \mu$.
\newline
In region $r > a$, we have 
$R_{1}\approx r-a\cos \theta $ and $%
R_{2}\approx r+a\cos \theta $. So, 
in terms of new coordinates $\mu$ and $\lambda$, the equation (\ref{a0}) turns into 
\begin{equation}
\begin{split}
(\mu ^{2}-a^{2})\frac{\partial ^{2}R(\mu ,\lambda )}{\partial \mu ^{2}}+2\mu 
\frac{\partial R(\mu ,\lambda )}{\partial \mu }+(a^{2}-\lambda ^{2})\frac{%
\partial ^{2}R(\mu ,\lambda )}{\partial \lambda ^{2}}-2\lambda \frac{%
\partial R(\mu ,\lambda )}{\partial \lambda }=& \\
c^{2}\left[  \epsilon (\mu
^{2}-\lambda ^{2})+3\mu n\right] R(\mu ,\lambda ).&
\label{pdeml}
\end{split}
\end{equation}%
This differential equation (\ref{pdeml}) separates into two ordinary second-order differential equations, given by 
\begin{eqnarray}
(\mu ^{2}-a^{2})\frac{d^{2}G(\mu )}{d\mu ^{2}}+2\mu \frac{dG(\mu )}{d\mu }%
-c^{2}( \epsilon \mu ^{2}+3\mu n+M^2)G(\mu ) &=&0  \label{sepmu} \\
(a^{2}-\lambda ^{2})\frac{d^{2}F(\lambda )}{d\lambda ^{2}}-2\lambda \frac{%
dF(\lambda )}{d\lambda }+c^{2}( \epsilon \lambda ^{2}+M^2)F(\lambda ) &=&0.
\label{seplambda}
\end{eqnarray}%
For $\mu \geq 2a$, introducing the new coordinate $0 \leq q\leq \tanh^{-1}(\frac{1}{2})$ related to $\mu$ by $\mu={\frac {a}{\tanh \left( q \right) }}$, the equation (\ref{sepmu}) changes to 
\begin{equation}
\frac{d^2G(q)}{dq^2}-\left( \frac{M^2c^2}{\sinh^2(q)}+\frac{\beta^2\cosh(q)}{\sinh^3(q)}
+\frac{\alpha^2\cosh^2(q)}{\sinh^4(q)}\right)G(q)=0
\label{Gq}
\end{equation}
where $\beta^2=3nc^2 a$, $\alpha^2=\epsilon c^2 a^2$.

The solutions to (\ref{Gq}) can be obtained as 
\begin{equation}
G_1(q)=g_1 q
{\mathcal{W}_{W}\left(-1/2\,{\frac {{\beta}^{2}}{\alpha}},\,1/2\,\sqrt {1+4\,{\gamma}^{2}},\,2\,{\frac {\alpha}{q}}\right)}
\label{G1q}
\end{equation}
where 
$\gamma^2={M}^{2}c^2+1/3\,{\alpha}^{2}$ and $g_1$ is a constant.
For $a  < \mu \leq 2a$, the solutions to (\ref{sepmu}) become
\begin{equation}
\begin{split}
&G_2(z)= {{e}^{-ca\sqrt {\epsilon}z}}{\cal H}_C \left( 4\,ca\sqrt {\epsilon
},0,0,6\,{c}^{2}an,-{c}^{2} \left( 3\,na+{M}^{2}+\epsilon\,{a}^{2}
 \right),\,{-\frac {z}{2}} \right) \times \\
 & (1+g_2\,
\int \!\frac{{e}^{2\,ca\sqrt {\epsilon}z}}{{z} \left( z+2\,
 \right) {\cal H}_C \left( 4\,ca\sqrt {\epsilon},0,0,6\,
{c}^{2}an,-{c}^{2} \left( 3\,na+{M}^{2}+\epsilon\,{a}^{2} \right),\,{-\frac {z}{2}} \right)^{2}}{dz}) 
\end{split}
\label{sepmu2zSC2}
\end{equation}
where $z=\frac{\mu}{a}-1$ and $g_2$ is a constant. We should note by choosing proper values for $g_1$ and $g_2$, two solutions (\ref{G1q}) and (\ref{sepmu2zSC2}) are $C^\infty$ continuous at $\mu=2a$.

For the second differential equation (\ref{seplambda}), the solutions are given by 
\begin{equation}
F(\lambda )=f_{cM}\mathcal{H}_{C}(0,-\frac{1}{2},0,-\frac{%
a^{2}c^{2} \epsilon }{4},\frac{1}{4}-\frac{M^{2}c^{2}}{4},\frac{\lambda
^{2}}{a^{2}})+f_{cM}^{\prime }\mathcal{H}_{C}(0,\frac{1}{2},0,-\frac{%
a^{2}c^{2} \epsilon }{4},\frac{1}{4}-\frac{M^{2}c^{2}}{4},\frac{\lambda
^{2}}{a^{2}})\lambda  \label{mzone2}
\end{equation}
where $f_{cM}$, and $f_{cM}^{\prime }$ are constants.

\begin{figure}[!ht]
\centering           
\begin{minipage}[c]{.6\textwidth}
        \centering
        \includegraphics[width=11cm]{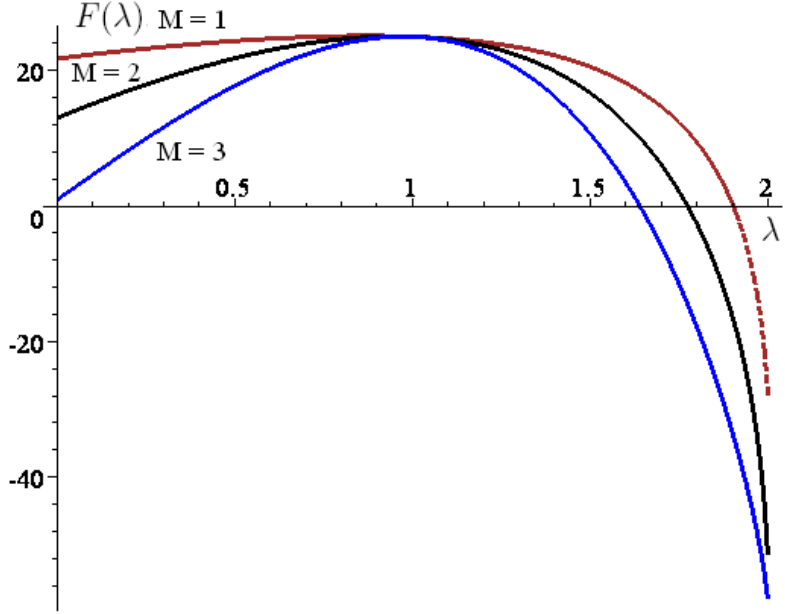}
    \end{minipage}
\caption{Numerical solutions to equation (\ref{seplambda}). }
\label{InterMFN}
\end{figure}

For completeness, we also numerically solve the equation (\ref{seplambda}) and the results are illustrated in Figure (\ref{InterMFN}).
\newline

As the final result, the most general solution for the M2-brane metric function in region $r>a$, is given by:
\begin{equation}
H(y,r,\theta)=1+Q_{M_2}\int_0^\infty dC\int_0^\infty dM \frac{J_1(cy)}{y}  G_t(\mu) F(\lambda)
\label{threechargeH}
\end{equation}
where $G_t(\mu)=G_1(\tanh^{-1}(\frac{a}{\mu}))\theta(\frac{\mu}{a}-2)+G_2(\frac{\mu}{a}-1)\theta(2-\frac{\mu}{a})$. 

Dimensional reduction of M2-brane metric (\ref{ds11m2}) with the metric
functions (\ref{Hyrsep}) along the coordinate $\psi $ of
the metric (\ref{dsGH}) gives type IIA supergravity metric 
\begin{eqnarray}
ds_{10}^{2} &=&H^{-1/2}(y,r,\theta)V^{-1/2}(r,\theta)\left(
-dt^{2}+dx_{1}^{2}+dx_{2}^{2}\right) +  \notag \\
&+&H^{1/2}(y,r,\theta)V^{-1/2}(r,\theta)\left( dy^{2}+y^{2}d\Omega
_{3}^{2}\right)+  \notag \\
&+&H^{1/2}(y,r,\theta)V^{1/2}(r,\theta)(dr^{2}+r^{2}d\Omega
_{2}^{2})  \label{ds10TN4}
\end{eqnarray}
which describes a localized D2-brane at $y=r=0$ along the world-volume of
D6-brane.
The only non-vanishing NSNS
field in ten dimensions is given by
\begin{equation}
\Phi =\frac{3}{4}\ln \left\{ \frac{H^{1/3}(y,r,\theta)}{%
V(r,\theta)}\right\} 
\end{equation}
while the Ramond-Ramond (RR) fields are
\begin{eqnarray}
C_{\phi } &=&\omega(r,\theta) \\
A_{tx_{1}x_{2}} &=&\frac{1}{H(y,r,\theta)}.
\end{eqnarray}
The intersecting configuration is BPS since it has been obtained by
compactification along a transverse direction from the BPS membrane solution
with harmonic metric function (\ref{Hyrsep}) \cite{forbps}.%

\section{M5-Brane Solutions, M2-Brane Solutions With Two Transverse Gibbons-Hawking Spaces and the Number of Preserved Supersymmetries}
\label{sec:M2mix}
To embed the Gibbons-Hawking space into the eleven dimensional M5-brane metric, we consider
\begin{eqnarray}
ds_{11}^{2}&=&H(y,r,\theta)^{-1/3}\left(
-dt^{2}+dx_{1}^{2}+dx_{2}^{2}+dx_{3}^{2}+dx_{4}^{2}+dx_{5}^{2}\right) + 
\notag \\
&+&H(y,r,\theta)^{2/3}\left( dy^{2}+ds_{GH}^{2}\right)  \label{ds11m5p}
\end{eqnarray}
with field strength components 
\begin{eqnarray}
F_{\psi \phi r y}&=&\frac{\alpha }{2}\sin (\theta )\frac{\partial H}{
\partial \theta}  \notag \\
F_{\psi \phi \theta y}&=&-\frac{\alpha }{2}r^2\sin (\theta )\frac{ \partial H%
}{\partial r}  \notag \\
F_{\psi \phi \theta r}&=&\frac{\alpha }{2}r^2\sin (\theta )V(r,\theta)\frac{
\partial H}{\partial y}.  \label{FScompo}
\end{eqnarray}
The M5-brane corresponds to $\alpha =+1$; while the $\alpha =-1$
case corresponds to an anti-M5 brane.

The metric (\ref{ds11m5p}) along with (\ref{FScompo}) are solutions to the supergravity equations 
provided $H\left( y,r,\theta\right) $ satisfies the
differential equation 
\begin{align}  \label{HeqforM5}
& & 2r\frac{\sin\theta}{V(r,\theta)}\frac{\partial H}{\partial r}+%
\frac{\cos\theta}{V(r,\theta)}\frac{\partial H}{\partial\theta}%
+r^2\sin\theta\frac{\partial^{2}H}{\partial y^{2}} +\frac{\sin\theta}{%
V(r,\theta)}\{\frac{\partial^{2}H}{\partial\theta^{2}}+r^2\frac{%
\partial^{2}H}{\partial r^{2}}\}=0.  \notag \\
& & 
\end{align}
Upon substituting 
$ H(y,r,\theta)=1+Q_{M5}Y(y)R(r,\theta)
$, where $Q_{M5}$ is the charge on the M5-brane, 
the equation (\ref{HeqforM5})
straightforwardly separates. The solution to the
differential equation for $Y(y)$ is a sine-harmonic function and the differential equation for $R(r,\theta)$ is the same equation as (\ref%
{a0}). Hence the most general M5-brane function, corresponding to embedded
Gibbons-Hawking space with $k=3$ is given by 
\begin{equation}
H(y,r,\theta)=1+Q_{M5}\int_0^\infty dc \int_0^\infty d{M}
\cos(cy+c')\times  R(r,\theta)
\label{M5sol1}
\end{equation}
where $c'$ is a constant, $R(r,\theta)$ is given by (\ref{nearchagefunapp}) for  region $r<a$ and (\ref{threechargeH}) for region $r>a$, respectively.  
Reducing (\ref{ds11m5p}) to ten dimensions gives
the following NSNS dilaton 
\begin{equation}
\Phi =\frac{3}{4}\ln \left\{ \frac{H^{2/3}(y,r,\theta)}{V(r,\theta)}%
\right\} .  \label{dilTN}
\end{equation}
The NSNS field strength of the two-form associated with the NS5-brane, is
given by 
\begin{equation}
\mathcal{H}_{(3)}={F_{\phi y r\psi }}d\phi \wedge dy \wedge dr+{F_{\phi y
\theta \psi}}d\phi \wedge dy \wedge d\theta +{F_{\phi r \theta \psi }}d\phi
\wedge dr \wedge d\theta  \label{tnH3}
\end{equation}
where the different components of 4-form $F$, are given by ( \ref{FScompo}).
The RR fields are 
\begin{eqnarray}
C_{(1)} &=&\omega(r,\theta )  \label{tnRR} \\
\mathcal{A}_{\alpha \beta \gamma } &=&0  \label{tnA3}
\end{eqnarray}
where $C_{\alpha }$ is the field associated with the D6-brane, and the
metric in ten dimensions is given by: 
\begin{eqnarray}
ds_{10}^{2} &=&V^{-1/2}(r,\theta)\left(
-dt^{2}+dx_{1}^{2}+dx_{2}^{2}+dx_{3}^{2}+dx_{4}^{2}+dx_{5}^{2}\right)
+H(y,r,\theta) V^{-1/2}(r,\theta)dy^{2}+  \notag \\
&+&H(y,r,\theta)V^{1/2}(r,\theta)\left( dr^{2}+r^{2}d\Omega
_{2}^{2}\right).  \label{tng10}
\end{eqnarray}
From (\ref{tnH3}), (\ref{tnRR}), (\ref{tnA3}) and the metric (\ref{tng10}),
we can see the above ten dimensional metric is an NS5$\perp $D6(5) brane
solution. We have explicitly checked the BPS 10-dimensional metric (\ref%
{tng10}), with the other fields (the dilaton (\ref{dilTN}), the 1-form field
(\ref{tnRR}), and the NSNS field strength (\ref{tnH3})) make a solution to
the 10-dimensional supergravity equations of motion. 
In addition to the solutions presented in sections 3 and 4, we can also embed two four dimensional Gibbons-Hawking spaces into the
eleven dimensional membrane metric. Here we consider the embedding of two
metrics of the form (\ref%
{dsGH}) with $k=3$. The M-brane metric is 
\begin{equation}
ds_{11}^{2} =H(y,\alpha,r,\theta)^{-2/3}\left(
-dt^{2}+dx_{1}^{2}+dx_{2}^{2}\right) +H(y,\alpha,r,\theta)^{1/3}\left(
ds_{GH(1)}^{2}+ds_{GH(2)}^{2}\right)  \label{tn4xtn4mtrc}
\end{equation}
where $ds_{GH(i)},\, i=1,2$ are two copies of the metric (\ref{dsGH}) with
coordinates $(r,\theta,\phi,\psi)$ and $(y,\alpha,\beta,\gamma)$. The
non-vanishing components of four-form field are 
\begin{equation}
F_{tx_{1}x_{2}x}=-\frac{1}{2H^{2}}\frac{\partial H(y,\alpha,r,\theta)}{%
\partial x}  \label{tn4xtn4F012i}
\end{equation}
where $x=r,\theta,y,\alpha$. The metric (\ref{tn4xtn4mtrc}) and four-form
field (\ref{tn4xtn4F012i}) satisfy the eleven dimensional equations of
motion if 
\begin{eqnarray}
& &2ry\sin(\alpha)\sin(\theta)\{V(r,\theta)y\frac{\partial H}{%
\partial r}+V(y,\alpha)r\frac{\partial H}{\partial y}\}+  \notag \\
&+& \sin(\alpha)y^2\cos(\theta)V(r,\theta)\frac{\partial H}{%
\partial \theta}+r^2\sin(\theta)\cos(\alpha)V(y,\alpha)\frac{%
\partial H}{\partial \alpha}+  \notag \\
&+& r^2\sin(\alpha)y^2\sin(\theta)\{V(r,\theta)\frac{\partial^2 H}{%
\partial r^2}+V(y,\alpha)\frac{\partial^2 H}{\partial y^2}\}+ 
\notag \\
&+&\sin(\theta)\sin(\alpha)\{r^2V(y,\alpha)\frac{\partial^2 H}{%
\partial \alpha^2}+y^2V(r,\theta)\frac{\partial^2 H}{\partial
\theta^2}\}=0  \label{deqtnxtn}
\end{eqnarray}
where $V(y,\alpha)=\epsilon+\hat n \{ \frac{1}{y}+\frac{1}{\sqrt{%
y^2+b^2+2by\cos(\alpha)}}+\frac{1}{\sqrt{%
y^2+b^2-2by\cos(\alpha)}}\}$. The equation (\ref{deqtnxtn}) is separable if we
set $H(y,\alpha,r,\theta)=1+Q_{M2}R_{1}(y,\alpha)R_{2}(r,\theta)$. This
gives two equations 
\begin{equation}
2x_i\frac{\partial R_i}{\partial x_i}+x_i^{2}\frac{\partial ^{2}R_i}{%
\partial x_i^{2}}+\frac{\cos y_i}{\sin y_i}\frac{\partial R_i}{\partial y_i}+%
\frac{\partial^{2}R_i}{\partial^{2} y_i}=u_ic^{2}x_i^{2}V(x_i,y_i)R_i  \label{tn4xtn4deqRi}
\end{equation}
where $(x_1,y_1)=(y,\alpha)$ and $(x_2,y_2)=(r,\theta)$. There is no
summation on index $i$ and $u_1=+1,\,u_2=-1$, in equation (\ref{tn4xtn4deqRi}%
). We already know the solutions to the two differential equations (\ref%
{tn4xtn4deqRi}) as given by 
(\ref{nearchagefunapp}) for region $r<a$ and (\ref{threechargeH}) region $r>a$.  
So the most
general solution to (\ref{deqtnxtn}) is 
\begin{equation}
H(y,\alpha,r,\theta)=1+Q_{M2}\int_{0}^{\infty}dc \int_0^\infty dM
\int_0^\infty d\tilde{M} R(y,\alpha)\tilde{R}(r,\theta).  \label{HTN4TN4}
\end{equation}

We can choose to compactify down to ten dimensions by compactifying on
either $\psi$ or $\gamma$ coordinates. In the first case, we find the type
IIA string theory with the only non-vanishing NSNS field 
as
\begin{equation}
\Phi =\frac{3}{4}\ln \left( \frac{H^{1/3}}{V(r,\theta)}\right)
\label{tnXtnNSNS}
\end{equation}
and RR fields 
\begin{eqnarray}
C_{\phi} & = & \omega(r,\theta)  \label{oh2} \\
A_{tx_{1}x_{2}} & = & H(y,\alpha,r,\theta)^{-1}.  \label{tnXtnRR}
\end{eqnarray}
The metric is given by 
\begin{eqnarray}
ds_{10}^{2} &=&H(y,\alpha,r,\theta)^{-1/2}{V(r,\theta)}%
^{-1/2}\left( -dt^{2}+dx_{1}^{2}+dx_{2}^{2}\right) +  \notag \\
&+&H(y,\alpha,r,\theta)^{1/2}{V(r,\theta)}^{-1/2}\left(
ds_{GH(1)}^{2}\right) +  \notag \\
&+&H(y,\alpha,r,\theta)^{1/2}{V(r,\theta)}^{1/2}\left(
dr^{2}+r^{2}\left( d\theta^{2}+\sin ^{2}(\theta)d\phi ^{2}\right) \right).
\label{ds10tnXtn}
\end{eqnarray}
In the latter case, the type IIA fields and metric are in the same form as (\ref{tnXtnNSNS}), (\ref{oh2}), (\ref{tnXtnRR}) and (\ref{ds10tnXtn}), just
by replacements $(r,\theta,\phi,\psi) \Leftrightarrow
(y,\alpha,\beta,\gamma) $. In either cases, we get a fully localized D2/D6
brane system. We can further reduce the metric (\ref{ds10tnXtn}) along the $%
\gamma$ direction of the first Gibbons-Hawking space. However the result of
this compactification is not the same as the reduction of the M-theory
solution (\ref{tn4xtn4mtrc}) over a torus, which is compactified type IIB
theory. The reason is that to get the compactified type IIB theory, we
should compactify the T-dual of the IIA metric (\ref{ds10tnXtn}) over a
circle, and not directly compactify the 10D IIA metric (\ref{ds10tnXtn})
along the $\gamma$ direction. We note also an interesting result in reducing
the 11D metric (\ref{tn4xtn4mtrc}) along the $\psi$ (or $\gamma$) direction
of the $GH(1)$ (or $GH(2)$) in large radial coordinates. As $y$ (or $r$) $%
\rightarrow \infty $\ the transverse geometry in (\ref{tn4xtn4mtrc}) locally
approaches $\mathbb{R}^{3}\otimes S^{1}\otimes GH(2)$ (or $GH(1)\otimes 
\mathbb{R}^{3}\otimes S^{1}$). Hence the reduced theory, obtained by
compactification over the circle of the Gibbons-Hawking, is IIA. Then by
T-dualization of this theory (on the remaining $S^{1}$ of the transverse
geometry), we find a type IIB theory which describes the D5 defects. The
solutions (\ref{tn4xtn4mtrc}) (with $\epsilon=0$ or $\epsilon\neq 0$) are
BPS and also preserve 1/4 of the supersymmetry similar to all other solutions in this paper.
Generically a
configuration of $n$ intersecting branes preserves $\frac{1}{2^n}$ of the
supersymmetry. In general, the Killing spinors are projected out by product
of Gamma matrices with indices tangent to each brane. If all the projections
are independent, then $\frac{1}{2^n}$-rule can give the right number of
preserved supersymmetries. On the other hand, if the projections are not
independent then $\frac{1}{2^n}$-rule can't be trusted. There are some
important brane configurations when the number of preserved supersymmetries
is more than that by $\frac{1}{2^n}$-rule \cite{ex,ex2}.
The number of non-trivial
solutions to the Killing spinor equation 
\begin{equation}
\partial_{M} \varepsilon+\frac{1}{4}\omega_{abM}\Gamma^{ab} \varepsilon+\frac{1%
} {144}\Gamma_{M}^{\phantom{m}npqr}F_{npqr} \varepsilon-\frac{1}{18}\Gamma
^{pqr}F_{mpqr} \varepsilon=0  \label{killingspinoreq2}
\end{equation}
determine the amount of supersymmetry of the solution where the indices $%
M,N,P,...$ are eleven dimensional world indices and $a,b,...$ are eleven
dimensional non-coordinate tangent space indices. In \cite{Rahim}, the authors presented the calculations explicitly to find how many supersymmetries are preserved for M2 and M5 brane solutions where the transverse space contains at least one Gibbons-Hawking of $k=2$ geometry . The explicit calculation enjoys the independence on explicit form of metric function $V(r,\theta)$ and $\omega (r,\theta)$. Hence we conclude all our solutions presented in previous sections preserve eight supersymmetries.
In fact, half of the
supersymmetry is removed by the projection operator that is due to the
presence of the brane, and another half is removed due to the self-dual
nature of the Gibbons-Hawking metric with $k=3$ or in general for any value of $k$. 

\section{Decoupling Limits of Solutions}

\label{sec:dec}

In this section we consider the decoupling limits of the solutions in different regions which are presented in sections 3,4,5 and appendix A. Since the specifics of calculating the decoupling limit
are shown in detail elsewhere (see for example \cite{DecouplingLim}), so we
will only provide a brief outline here. The process is the same for all
cases, so we will also only provide specific examples of a few of the
solutions in different regions that presented in sections \ref{sec:Nnutcharge},  \ref{sec:solutionsMiddle}, \ref{sec:M2mix} and appendix A.

At low energies, the dynamics of the D2 brane decouple from the bulk, with
the region close to the D6 brane corresponding to a range of energy scales
governed by the IR fixed point \cite{DecouplingLim1}. For D2 branes
localized on D6 branes, this corresponds in the field theory to a vanishing
mass for the fundamental hyper-multiplets. Near the D2 brane horizon ($H\gg
1 $), the field theory limit is given by 
$g_{YM2}^{2}=g_{s}\ell _{s}^{-1}=\text{fixed}$.
In this limit the gauge couplings in the bulk go to zero, so the dynamics
decouple there. In each of our cases above, we scale the coordinates $y$ and 
$r$ given by
$ y=Y\ell _{s}^{2} $ and $r=U\ell _{s}^{2}$ respectively,
such that $Y$ and $U$ are fixed. We
note that this will change the harmonic function of the D6 brane in the
Gibbons-Hawking case $(k=3)$ to the following  
\begin{equation}
V(U,\theta)=\epsilon+g_{YM2}^{2}N_{6}\{\frac{1}{U}+\frac{1}{\sqrt{%
U^2+A^2+2AU\cos\theta}}+\frac{1}{\sqrt{%
U^2+A^2-2AU\cos\theta}}\}  \label{F2}
\end{equation}
where we rescale $a$ to $a=A\ell _{s}^{2}$ and generalize to the case of $%
N_{6}$ D6 branes. We also recall that to avoid any conical
singularity, we should have $n_1=n_2=n_3=n$, hence the asymptotic radius of the
11th dimension is $R_{\infty }=n=g_{s}\ell _{s} $. We show that the metric function $H(y,r,\theta)$ always scales
as $H(Y,U,\theta)=\ell _{s}^{-4}h(Y,U,\theta)$ if the coefficients 
of solutions in different regions, obey some specific scaling. The scaling
behavior of $H(Y,U,\theta)$ causes then the D2-brane to warp the ALE region
and the asymptotically flat region of the D6-brane geometry. As the first example, we consider the solutions given by (\ref{NearNutChargef}) and (\ref{NearNutChargeg}) and calculate $h(Y,U,\theta)$. 
After scaling, we get 
\begin{eqnarray}
h(Y,U,\theta)&=&32\pi^2N_2g_{YM}^4\int _0 ^\infty dC \int _0 ^ \infty {\cal M}d{\cal M}\frac{J_1(CY)}{Y}\nonumber\\
&\times&\big(\frac{F_0}{U}\mathcal{W}_{M}(-\frac{C{\cal N}}{2\sqrt{\epsilon +\tilde{A}}},\frac{%
\sqrt{1+4{\cal M}^2C^{2}}}{2},2C\sqrt{\epsilon + \tilde{A}}U)\nonumber\\
&\times& e^{- \tilde{\beta} \zeta} 
{\cal F}(\Xi ,{1-\Xi },{1},\frac{1}{2}(1-\zeta)
)\big(G_1+G_{2}\int \frac{d\zeta}{(\zeta^2-1) {\cal F}(\Xi ,{1-\Xi },{1},\frac{1}{2}(1-\zeta)
)^2 } \big)
\nonumber\\
&
\label{Hint2}
\end{eqnarray} 
where we scale the coefficients to $F_0=f_0\ell_s^4$, $G_1=g_1\ell_s^6$ and $G_2=g_2\ell_s^6$ as well as separation constants to $\tilde{\beta}=B {\cal N} C^2 A $, $C=c\ell _{s}^{2}$ and $M={\cal M}\ell_{s}^{2}$. Moreover ${\cal N}=\frac{n}{\ell _s^2}$, $a=\ell_s^2 A$ and $\Xi=\frac{1}{2}+\frac{\sqrt{1+4{
\cal M}^2C^{2}}}{2}$ or $\Xi=-\frac{1}{2}-\frac{\sqrt{1+4{\cal M}^2C^{2}-4 \tilde{\beta}^2}}{2}$ and $\tilde{A}=(\sum^{N_1}_{k=1}\frac{1}{k}+\sum^{N_2}_{k=1}\frac{1}{k})\times \frac{{\cal N}}{A} $. We should note in (\ref{Hint2}) we use $\ell
_{p}=g_{s}^{1/3}\ell _{s}$ to rewrite $Q_{M2}=32\pi ^{2}N_{2}\ell _{p}^{6}$
in terms of $\ell_s$ given by $Q_{M2}=32\pi ^{2}N_{2}g_{YM2}^{4}\ell
_{s}^{8}$.
For the second example, we consider solutions given by (\ref{threechargeH}). 
The rescaled metric function $h(Y,U,\theta)$ read as
\begin{equation}
h(Y,U,\theta)=32\pi ^{2}N_{2}g_{YM2}^{4}\int _0 ^\infty dC \int _0 ^\infty {\cal M}d{\cal M} \frac{J_1(CY)}{Y}G_t(\Psi)F(\Lambda).
\label{hmiddle}
\end{equation}
In (\ref{hmiddle}), $G_t(\Psi)=G_1(\tanh^{-1}(\frac{A}{\Psi})\Theta(\frac{\Psi}{A}-2)+G_2(Z)\Theta(2-\frac{\Psi}{A})$ in terms of scaled coordinate $\Psi=\frac{\mu}{\ell_{s}^2}$, where   
\begin{equation}
\begin{split}
&G_2(Z)= {{e}^{-CA\sqrt {\epsilon}Z}}{\cal H}_C \left( 4\,CA\sqrt {\epsilon
},0,0,6\,{C}^{2}AN,-{C}^{2} \left( 3\,NA+{{\cal M}}^{2}+\epsilon\,{A}^{2}
 \right),\,{-\frac {Z}{2}} \right) \times \\
 & \,(1+G_2
\int \!\frac{{e}^{2\,CA\sqrt {\epsilon}Z}}{{Z} \left( Z+2\,
 \right) {\cal H}_C \left( 4\,CA\sqrt {\epsilon},0,0,6\,
{C}^{2}AN,-{C}^{2} \left( 3\,NA+{{\cal M}}^{2}+\epsilon\,{A}^{2} \right),\,{-\frac {Z}{2}} \right)^{2}}{dZ}).
\end{split}
\label{GZ}
\end{equation}
The scaled quantities in (\ref{GZ}) are $a=A\ell_s^2$, $n=N \ell _s^2$.  $G_2=g_2$ and $Z$ is given by $Z=\frac{\Psi}{A}-1$. The other part of integrand in (\ref{hmiddle}) is
\begin{equation}
F(\Lambda )=F_{C{\cal M}}\mathcal{H}_{C}(0,-\frac{1}{2},0,-\frac{%
A^{2}C^{2} \epsilon }{4},\frac{1}{4}-\frac{{\cal M}^{2}C^{2}}{4},\frac{\Lambda
^{2}}{A^{2}})+F_{C {\cal M}}^{\prime }\mathcal{H}_{C}(0,\frac{1}{2},0,-\frac{%
A^{2}C^{2} \epsilon }{4},\frac{1}{4}-\frac{{\cal M}^{2}C^{2}}{4},\frac{\Lambda
^{2}}{A^{2}})\Lambda  \label{FLam}
\end{equation}
where $\lambda=\Lambda \ell _s^2$,  
$F_{C{\cal M}}=f_{cM}\ell _s^6$ and 
$F_{C{\cal M}}^\prime=f_{cM}^\prime\ell _s^8$.

In all other cases we can show we have the same scaling behavior as $h(Y,U,\theta)=\ell _{s}^{4}H(Y,U,\theta)$. 
In any case, the respective ten-dimensional supersymmetric metric (\ref{ds10TN4}) scales
as 
\begin{eqnarray}
\frac{ds_{10}^{2}}{\ell
_{s}^{2}} &=&h^{-1/2}(Y,U,\theta)V^{-1/2}(U,\theta)\left(
-dt^{2}+dx_{1}^{2}+dx_{2}^{2}\right) +  \notag \\
&+&h^{1/2}(Y,U,\theta)V^{-1/2}(U,\theta)\{\left( dY^{2}+Y^{2}d\Omega
_{3}^{2}\right)+ V(U,\theta)(dU^{2}+U^{2}d\Omega
_{2}^{2})\} \label{m10}
\end{eqnarray}
that shows only one overall normalization factor of $\ell _{s}^{2}$ in
the metric (\ref{m10}). This is the expected result for a solution that is a
supergravity dual of a QFT.  We now consider an analysis of the decoupling limits of M5-brane solution
given by metric function (\ref{M5sol1}).

At low energies, the dynamics of IIA NS5-branes will decouple from the bulk 
\cite{DecouplingLim2}. Near the NS5-brane horizon ($H>>1$), we are
interested in the behavior of the NS5-branes in the limit where string
coupling vanishes 
$ g_{s}\rightarrow 0$
while 
$\ell _{s}=$fixed.
In these limits, we rescale the radial coordinates by $Y=\frac{y}{g_{s}\ell _{s}^{2}}$ and $U=\frac{r}{g_{s}\ell _{s}^{2}}$ such that they can be
kept fixed. This causes the Gibbons-Hawking harmonic function of the D6-brane solution (\ref{tng10}), change to 
\begin{equation}
V(U,\theta)=\epsilon+\frac{N_6}{\ell _{s}}\{\frac{1}{U}+
\frac{1}{\sqrt{U^2+A^2+2AU\cos\theta}}
+\frac{1}{\sqrt{U^2+A^2-2AU\cos\theta}}
\}
\end{equation}
where we generalize to $N_{6}$ D6-branes and rescale $a=A\ell _{s}^{2}g_{s}$.

Similar to what we did for M2-branes, we easily can show the harmonic functions for M5-branes (\ref{M5sol1}), rescale
according to $H(Y,U,\theta)=g_{s}^{-2}h(Y,U,\theta)$ such that
$h(Y,U,\theta)$ doesn't have any $g_{s}$
dependence \cite{Rahim}.

As a result, in decoupling limit, the ten-dimensional metric (\ref{tng10}) becomes, 
\begin{eqnarray}
ds_{10}^{2} &=&V^{-1/2}(U,\theta)\left(
-dt^{2}+dx_{1}^{2}+dx_{2}^{2}+dx_{3}^{2}+dx_{4}^{2}+dx_{5}^{2}\right)
\nonumber \\
&+&\ell_s^4\{h(Y,U,\theta) V^{-1/2}(U,\theta)dY^{2}
+h(Y,U,\theta)V^{1/2}(U,\theta)\left( dU^{2}+U^{2}d\Omega
_{2}^{2}\right)\}.  \label{tng100}
\end{eqnarray}

In the limit of vanishing $g_{s}$\ with fixed $\ell_{s}$, 
the decoupled free theory on NS5-branes should be
a little string theory \cite{shiraz} (i.e. a 6-dimensional non-gravitational
theory in which modes on the 5-brane interact amongst themselves, decoupled
from the bulk). We note that our NS5/D6 system is obtained from M5-branes by
compactification on a circle of self-dual transverse geometry. Hence the IIA
solution has T-duality with respect to this circle. The little string theory
inherits the same T-duality from IIA string theory, since taking the limit
of vanishing string coupling commutes with T-duality. Moreover T-duality
exists even for toroidally compactified little string theory. In this case,
the duality is given by an $O(d,d,\mathbb{Z})$ symmetry where $d$\ \ is the
dimension of the compactified toroid. These are indications that the little
string theory is non-local at the energy scale $l_{s}^{-1}$ and in
particular in the compactified theory, the energy-momentum tensor can't be
defined uniquely \cite{aha}.

As the last case, we consider the analysis of the decoupling limits of the
IIB solution that can be obtained by T-dualizing the compactified M5-brane
solution (\ref{ds11m5p}). The type IIA NS5$\perp $ D6(5) configuration is
given by the metric (\ref{tng10}) and fields (\ref{dilTN}), (\ref{tnH3}), (%
\ref{tnRR}) and (\ref{tnA3}).

We apply the T-duality \cite{Cascaless} in the $x_{1}-$direction of the
metric (\ref{tng10}), that yields gives the IIB dilaton field 
\begin{equation}
\widetilde{\Phi }=\frac{1}{2}\ln \frac{H}{\tilde{f}}  \label{IBdilaton}
\end{equation}
the 10D type IIB metric, as 
\begin{eqnarray}
\widehat{ds}_{10}^{2} &=&V^{-1/2}(r,\theta)\left(
-dt^{2}+V(r,%
\theta)dx_{1}^{2}+dx_{2}^{2}+dx_{3}^{2}+dx_{4}^{2}+dx_{5}^{2}\right) + 
\notag \\
&+&H(y,r,\theta) V^{-1/2}(r,\theta)dy^{2}+
H(y,r,\theta)V^{1/2}(r,\theta)\left( dr^{2}+r^{2}d\Omega
_{2}^{2}\right).  \label{IIBmetric}
\end{eqnarray}
The metric (\ref{IIBmetric}) describes a IIB NS5$\perp $D5(4) brane
configuration (along with the dualized dilaton, NSNS and RR fields).

At low energies, the dynamics of IIB NS5-branes will decouple from the bulk.
Near the NS5-brane horizon ($H>>1$), the field theory limit is given by 
\begin{equation}
g_{YM5}=\ell _{s}=\text{ fixed}  \label{gym}
\end{equation}
The harmonic function of the D5-brane is 
\begin{equation}
V(r,\theta)=\epsilon+\frac{N_5}{g_{YM5}}\{\frac{1}{U}+\frac{1}{%
\sqrt{U^2+A^2+2AU\cos\theta}} + \frac{1}{\sqrt{U^2+A^2-2AU\cos\theta}} \}
\end{equation}
where ${N}_{5}$\ is the number of D5-branes.

The harmonic function of the NS5$\perp $D5 system (\ref{IIBmetric}),
rescales according to $H(Y,U,\theta)=g_{s}^{-2}{h}(Y,U,\theta)$, and the ten-dimensional metric (\ref{IIBmetric}), in the decoupling limit, becomes 
\begin{eqnarray}
\widetilde{ds}_{10}^{2} &=&V^{-1/2}(U,\theta)\left(
-dt^{2}+V(U,%
\theta)dx_{1}^{2}+dx_{2}^{2}+dx_{3}^{2}+dx_{4}^{2}+dx_{5}^{2}\right)+  \notag
\\
&+&g_{YM5}^2h(Y,U,\theta) \{V^{-1/2}(U,\theta)dY^{2}+
+V^{1/2}(U,\theta)\left( dU^{2}+U^{2}d\Omega _{2}^{2}\right)\}.
\end{eqnarray}

The decoupling limit illustrates that the decoupled theory in the low energy
limit is super Yang-Mills theory with $g_{YM}=\ell _{s}.$\ In the limit of
vanishing $g_{s}$\ with fixed $l_{s}$,\ the decoupled free theory on IIB
NS5-branes (which is equivalent to the limit $g_{s}\rightarrow \infty $\ of
decoupled S-dual of the IIB D5-branes) reduces to a IIB (1,1) little string
theory with eight supersymmetries.\ 

\section{Concluding Remarks}

The central thrust of this paper is the 
construction of
supergravity solutions for fully localized D2/D6 and NS5/D6 brane
intersections without restricting to the near core region of the D6 branes.
The metric functions of these solutions
is the dependence of the metric function 
depend to three (and four) transverse
coordinates. These solutions are new M2 and M5 brane metrics that are
presented in equations (\ref{NNa2}), (\ref{threechargeH}),
(\ref{M5sol1}) and (\ref{HTN4TN4}), which are the main results
of this paper. The common feature of all of these solutions is that the
brane function is a convolution of an decaying function with a damped
oscillating one. The metric functions vanish far from the M2 and M5 branes
and diverge near the brane cores.

Dimensional reduction of the M2 solutions to ten dimensions gives us
intersecting IIA D2/D6 configurations that preserve 1/4 of the
supersymmetry. For the M5 solutions, dimensional reduction yields IIA NS5/D6
brane systems overlapping in five directions. The latter solutions also
preserve 1/4 of the supersymmetry and in both cases the reduction yields
metrics with acceptable asymptotic behaviors.

We considered the decoupling limit of our solutions and found that D2 and
NS5 branes can decouple from the bulk, upon imposing proper scaling on some
of the coefficients in the integrands.

In the case of M2 brane solutions; when the D2 brane decouples from the
bulk, the theory on the brane is 3 dimensional $\mathcal{N}=4$ $SU($N$_{2})$
super Yang-Mills (with eight supersymmetries) coupled to N$_{6}$\ massless
hypermultiplets \cite{pelc}. This point is obtained from dual field theory
and since our solutions preserve the same amount of supersymmetry, a similar
dual field description should be attainable.

In the case of M5 brane solutions; the resulting theory on the NS5-brane in
the limit of vanishing string coupling with fixed string length is a little
string theory. In the standard case, the system of N$_{5}$\ NS5-branes
located at N$_{6}$\ D6-branes can be obtained by dimensional reduction of \ N%
$_{5}$N$_{6}$\ coinciding images of M5-branes in the flat transverse
geometry. In this case, the world-volume theory (the little string theory)
of the IIA NS5-branes, in the absence of D6-branes, is a non-local
non-gravitational six dimensional theory \cite{seiberg}. This theory has
(2,0) supersymmetry (four supercharges in the \textbf{4}\ representation of
Lorentz symmetry $Spin(5,1)$) and an R-symmetry $Spin(4)$ remnant of the
original ten dimensional Lorentz symmetry. The presence of the D6-branes
breaks the supersymmetry down to (1,0), with eight supersymmetries. Since we
found that some of our solutions preserve 1/4 of supersymmetry, we expect
that the theory on NS5-branes is a new little string theory. \ By
T-dualization of the 10D IIA theory along a direction parallel to the
world-volume of the IIA NS5, we find a IIB NS5$\perp $D5(4) system,
overlapping in four directions. The world-volume theory of the IIB
NS5-branes, in the absence of the D5-branes, is a little string theory with
(1,1) supersymmetry. The presence of the D5-brane, which has one transverse
direction relative to NS5 world-volume, breaks the supersymmetry down to
eight supersymmetries. This is in good agreement with the number of
supersymmetries in 10D IIB theory: T-duality preserves the number of
original IIA supersymmetries, which is eight. Moreover we conclude that the
new IIA and IIB little string theories are T-dual: the actual six
dimensional T-duality is the remnant of the original 10D T-duality after
toroidal compactification.

A useful application of the exact M-brane solutions in our paper is to
employ them as supergravity duals of the NS5 world-volume theories with
matter coming from the extra branes. More specifically, these solutions can
be used to compute some correlation functions and spectrum of fields of our
new little string theories.

In the standard case of $A_{k-1}$ (2,0) little string theory, there is an
eleven dimensional holographic dual space obtained by taking appropriate
small $g_s$ limit of an M-theory background corresponding to M5-branes with
a transverse circle and $k$ units of 4-form flux on $S^3 \otimes S^1$. In
this case, the supergravity approximation is valid for the (2,0) little
string theories at large $k$ and at energies well below the string scale.
The two point function of the energy-momentum tensor of the little string
theory can be computed from classical action of the supergravity evaluated
on the classical field solutions \cite{shiraz}.

Near the boundary of the above mentioned M-theory background, the string
coupling goes to zero and the curvatures are small. Hence it is possible to
compute the spectrum of fields exactly. In \cite{aha}, the full spectrum of
chiral fields in the little string theories was computed and the results are
exactly the same as the spectrum of the chiral fields in the low energy
limit of the little string theories. Moreover, the holographic dual theories
can be used for computation of some of the states in our little string
theories.

We conclude with a few comments about possible directions for future work.
Investigation of the different regions of the metric (\ref{ds11m5p}) or
alternatively the 10D string frame metric (\ref{tng100}) with a dilaton
for small and large Higgs
expectation value $U$\ would be interesting, as it could provide a means\
for finding a holographical dual relation to the new little string theory we
obtained. Moreover, the Penrose limit of the near-horizon geometry may be
useful for extracting information about the high energy spectrum of the dual
little string theory \cite{Gomiss}. The other open issue is the possibility
of the construction of a pp-wave spacetime which interpolates between the
different regions of the our new IIA NS5-branes. 

\vspace*{1cm}

{\Large Acknowledgments}

\vspace*{0.5cm}

This work was supported by the Natural Sciences and Engineering Research
Council of Canada.

\vspace*{1cm}

\appendix

{}

\section{Solutions around origin}

In this appendix, we present the solutions for M-brane metric functions in near region where $r<a$. In this region, we notice  
\begin{equation}
V(r,\theta)\approx \epsilon+\frac{n}{r}+\underbrace{\sum_{k=1}^{N_1}{\frac{n}{ka}}+\sum_{k=1}^{N_2}{\frac{n}{ka}}}_{A}+
\frac{nr \cos \theta}{a^2} \underbrace{\left[ \sum_{k=1}^{N_2}{\frac{1}{k^2}} -\sum_{k=1}^{N_1}{\frac{1}{k^2}} \right]}_{B}\label{A1}
\end{equation}
and the equation of motion (\ref{a0}) becomes 
\begin{equation}
 2r\frac{\partial R(r,\theta )}{\partial r}+r^{2}\frac{\partial
^{2}R(r,\theta )}{\partial r^{2}}+\frac{\cos \theta }{\sin \theta }\frac{%
\partial R(r,\theta )}{\partial \theta }+\frac{\partial ^{2}R(r,\theta )}{%
\partial ^{2}\theta }={c}^{2}{r}^{2} \left( \epsilon+A+{\frac {n}{r}}+{\frac {nBr\cos
  \theta }{{a}^{2}}} \right) R(r,\theta )  
\label{NnutchargeAAA}
\end{equation}
where we assume $B\neq 0$ ($N_1 \neq N_2$). If $B=0$, we should consider higher order terms in (\ref{A1}) which we will consider the case of $N_1=N_2=N_0$ later in this appendix.
We redefine  $R(r,\theta)$ as follows \\   
\begin{equation*}
R(r,\theta)=e^{\beta \cos \theta}\Psi(r,\theta)
\end{equation*}
where $\beta=\frac{naBc^2}{2}$. As we already know $(\frac{r}{a}<1)$, so the partial differential equation in terms of $\Psi(r,\theta)$ approximates to be 
\begin{equation}
\begin{split}
 2r\frac{\partial \Psi(r,\theta )}{\partial r}+r^{2}\frac{\partial
^{2}\Psi(r,\theta )}{\partial r^{2}}+&\left( \frac{\cos \theta }{\sin \theta }-2\beta \sin \theta \right) \frac{\partial \Psi(r,\theta )}{\partial \theta }+ \frac{\partial^{2}\Psi(r,\theta )}{\partial {\theta}^{2}} +{( \beta \sin \theta )}^2  \Psi(r,\theta )\\
-&{2\beta \cos \theta} \Psi(r,\theta )-c^2\left[ (\epsilon+A)r^2+nr \right] \Psi(r,\theta)=0.
\end{split}
\label{NnutchargeAAAA}
\end{equation}
The partial differential equation (\ref{NnutchargeAAAA}) separates into \begin{equation}
r^2 \frac{d^2{f(r)}}{d{r^2}} + 2r \frac{df(r)}{dr}-c^2\left[ (\epsilon+A)r^2+nr+M^2 \right]f(r)=0
\label{NearPN12}
\end{equation}
\begin{equation}
\frac{d^2{g(\theta)}}{d{{\theta}^2}} + \left( \frac{\cos \theta }{\sin \theta} - 2 \beta \sin \theta \right) \frac{dg(\theta)}{d \theta}+\left( M^2 c^2 - 2 \beta \cos \theta +{( \beta \sin \theta )}^2 \right) g(\theta)=0.
\label{NearPN121}
\end{equation}
Solution to (\ref{NearPN12}) is a Whittaker M function 
\begin{equation}
f(r)=\frac{f_0}{r}\mathcal{W}_{M}(-\frac{cn}{2\sqrt{\epsilon +A }},\frac{%
\sqrt{1+4M^2c^{2}}}{2},2c\sqrt{\epsilon+A }r).  \label{NearNutChargef}
\end{equation}
The solutions to (\ref{NearPN121}), in terms of coordinate $\zeta =\cos \theta $, are given by 
\begin{equation}
g(\zeta)=e^{-\beta \zeta} {\cal F}({\nu },1-\nu,1,\frac{1}{2}(1-\zeta
))\left[ g_1+g_{2}\int \frac{d\zeta}{(\zeta^2-1){\cal F}({\nu },1-\nu,1,\frac{1}{2}(1-\zeta
))^2}\right] \label{NearNutChargeg}
\end{equation}%
where ${\cal F}$ is the hypergeometric function and  $\nu=\frac{1}{2}+\frac{\sqrt{1+4M^2c^{2}}}{2}$. 
The solution can be expressed in the series forms as 
\begin{equation}
\begin{split}
g(\xi)=&C_1 \left( 1+\frac{2 \beta - M^2 c^2}{2}\xi + \cdots \right) + \\
&C_2 \left( ln(\xi) (1+\frac{ 2 \beta - M^2 c^2}{2}\xi + \cdots) + (\frac{1}{2}+ M^2 c^2 )\xi+ \cdots  \right)
\label{NearNutChargegS}
\end{split}
\end{equation}%
where  $\xi=1-\zeta$. 

As we mentioned before, if $N_1=N_2=N_0$, we should keep higher order terms in (\ref{A1}).
Starting from (\ref{a0}) and changing the coordinates to 
\begin{equation}
x=\cos (\theta ),\ z=\frac{r}{a}
\label{xrz}
\end{equation}
we get 
\begin{equation}
\begin{split}
z^{2}\frac{\partial ^{2}R(z,x)}{\partial z^{2}}+2z\frac{\partial R(z,x)}{%
\partial z}+&(1-x^{2})\frac{\partial ^{2}R(z,x)}{\partial x^{2}}-2x\frac{%
\partial R(z,x)}{\partial x} \\
&-\left[ c^{2}(a^{2}\epsilon
+2naA_0)z^{2}+nac^{2}z+naB_0c^{2}z^{4}(3x^{2}-1)\right] R(z,x)=0
\end{split}
\label{nearzonem}
\end{equation}%
where $A_0=\sum _{k=1}^{N_0} \frac{1}{k}$ and $B_0=\sum _{k=1}^{N_0} \frac{1}{k^2}$. 
To solve (\ref{nearzonem}), we introduce the function $\Omega(x,z)$ as follows
\begin{equation}
R(z,x)=e^{\beta x}\Omega \left( z,x \right)
\label{nearchagefunapp}
\end{equation}
where $\beta=\sqrt{3naB_0}c$. Hence 
the differential equation  (\ref{nearzonem}) 
in terms of $\Omega \left( z,x \right)$ becomes 
\begin{equation}
\begin{split}
& \left( 2\,\beta-2\,x-2\,{x}^{2}\beta \right) {\frac {\partial }{
\partial x}}\Omega \left( z,x \right) +2\,z{\frac {\partial }{
\partial z}}\Omega \left( z,x \right) + \left( 1-{x}^{2} \right) {
\frac {\partial ^{2}}{\partial {x}^{2}}}\Omega \left( z,x \right) +{z}
^{2}{\frac {\partial ^{2}}{\partial {z}^{2}}}\Omega \left( z,x
 \right) \\ 
&+ \left( {\beta}^{2}-2\,\beta\,x-{x}^{2}{\beta}^{2} \right) 
\Omega \left( z,x \right) + \left( na{c}^{2}B{z}^{4}-na{c}^{2}z+
 \left( -{c}^{2}{a}^{2}\epsilon-2\,{c}^{2}naA \right) {z}^{2} \right) 
\Omega \left( z,x \right) =0.
\end{split}
\label{nearchagenewfun}
\end{equation}
Separating the variables in $\Omega(z,x)$ by $\Omega (z,x)=\Upsilon (z)\Theta (x)$ and substituting into (\ref{nearchagenewfun}), we find two separated second order differential equations for $\Theta(x)$ and $\Upsilon(z)$, as follows  
\newline
\begin{equation}
\left( 1-{x}^{2} \right) {\frac {d^{2}}{d{x}^{2}}}\Theta \left( x \right) +2\, \left(  \left( 1-{x}^{2} \right) \beta-x \right) {\frac 
{d}{dx}}\Theta \left( x \right) - \left( 2\,x\beta+{\beta}^{2}{x}^{2}-
{M}^{2}{c}^{2}-\beta^2 \right) \Theta \left( x \right)=0
\label{nearzonemain2}
\end{equation}%
\begin{equation}
{z}^{2}{\frac {d^{2}}{d{z}^{2}}}{\Upsilon} \left( z \right) +2\,z{
\frac {d}{dz}}{\Upsilon} \left( z \right) + \left( -{M}^{2}{c}^{2}+
na{c}^{2}B_0{z}^{4}-na{c}^{2}z+ \left( -{c}^{2}{a}^{2}\epsilon-2\,{c}^{2
}naA_0 \right) {z}^{2} \right) {\Upsilon} \left( z \right) =0.
\label{nearzonemain1}
\end{equation}
The solutions to (\ref{nearzonemain2}) are given by (\ref{NearNutChargeg}) as $\Theta(x)=g(\zeta)\vert _{\zeta=x}$ while the solutions to (\ref{nearzonemain1}) can be written as
\begin{equation}
\Upsilon (z)={z}^{-\frac{\sqrt {4\,{M}^{
2}{c}^{2}+1}+1}{2}}
\Upsilon _1(z)+
{z}^{\frac{\sqrt {4\,{M}^{
2}{c}^{2}+1}-1}{2}}
\Upsilon _2(z)
\label{Upsilons}
\end{equation} 
where $\Upsilon _i (z),\,i=1,2$ are two independent polynomials of $z$.

\section{The Heun-C functions}

The Heun-C function $\mathcal{H}_{C}(\alpha,\beta,\gamma,\delta,\lambda,z)$
is the solution to the confluent Heun's differential equation \cite{Heuref} 
\begin{equation}
\mathcal{H}_{C}^{\prime \prime }+(\alpha+\frac{\beta+1}{z}+\frac{\gamma+1}{%
z-1})\mathcal{H}_{C}^{\prime }+ (\frac{\mu}{z}+\frac{\nu}{z-1})\mathcal{H}%
_{C}=0  \label{Heq}
\end{equation}
where $\mu=\frac{\alpha-\beta-\gamma+\alpha\beta-\beta\gamma}{2}-\lambda$
and $\nu=\frac{\alpha+\beta+\gamma+\alpha\beta+\beta\gamma}{2}%
+\delta+\lambda $. The equation (\ref{Heq}) has two regular singular points
at $z=0$ and $z=1 $ and one irregular singularity at $z=\infty$. The $%
\mathcal{H}_{C}$ function is regular around the regular singular point $z=0$
and is given by $\mathcal{H}_{C}=\Sigma_{n=0}^\infty
h_n(\alpha,\beta,\gamma,\delta,\lambda)z^n$, where $h_0=1$. The series is
convergent on the unit disk $\vert z\vert <1$ and the coefficients $h_n$ are
determined by the recurrence relation 
\begin{equation}
h_n=\Theta_nh_{n-1}+\Phi_nh_{n-2}
\end{equation}
where we set $h_{-1}=0$ and 
\begin{eqnarray}
\Theta_n&=&\frac{2n(n-1)+(1-2n)(\alpha-\beta-\gamma)+2\lambda-\alpha\beta+%
\beta\gamma}{2n(n+\beta)} \\
\Phi_n&=&\frac{\alpha(\beta+\gamma+2(n-1))+2\delta}{2n(n+\beta)}.
\end{eqnarray}

\section{Series expansion of some solutions}

The angular function (\ref{gg}) has a series expansion around $\xi=0$, given by

\begin{equation}
\begin{split}
g \left( \xi \right) =&C_{c,M}{\left[ 1-\frac{1}{2}{c^2(M^2+\tilde{m})}\xi+\cdots \right]}+ \\
&C'_{c,M}{\left[( 1-\frac{1}{2}{c^2(M^2+\tilde{m})}\xi+\cdots )\ln(\xi)+(\frac{1}{2}+c^2(M^2+\tilde{m}))\xi +\cdots \right]}
\end{split}
\label{Ngyseries}
\end{equation}

We notice an explicit logarithmically divergent behavior at $\xi=\theta=0$ as well as on figure \ref{NNNchargesF2}. The other divergent behavior of $g_2(\xi)$ at $\xi=2$ (in figure \ref{NNNchargesF2}) could be obtained easily by expansion of (\ref{gg}) around $\xi=2$.

\begin{figure}[!ht]
\centering           
\begin{minipage}[c]{.6\textwidth}
        \centering
        \includegraphics [width=11cm]{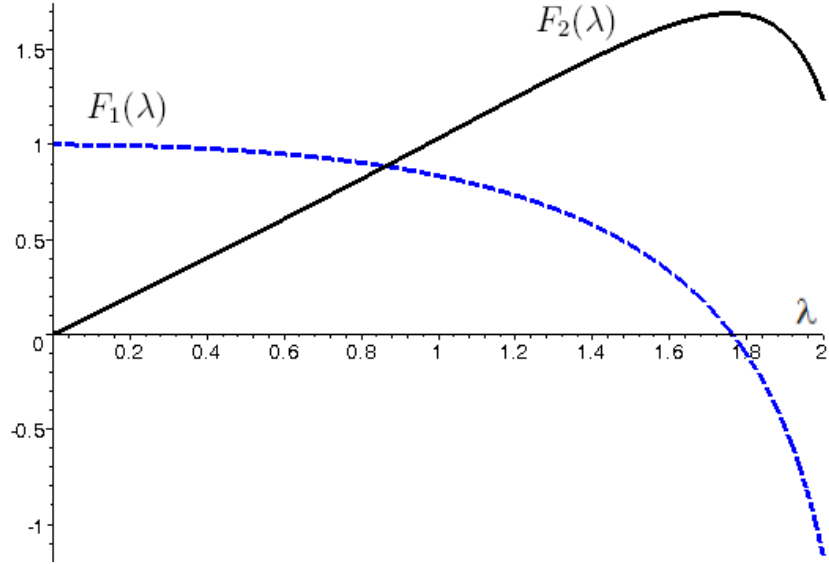} 
    \end{minipage}
\caption{$F_1(\lambda$) and $F_2(\lambda$) as given in (\ref{Gmu1}). }
\label{hofxii51}
\end{figure}

The series solution of (\ref{mzone2}) is given by 
\begin{equation}
\begin{split}
F(\lambda ) =&F_{I}\underbrace{\left[ 1-\frac{c^{2}M^{2}}{2a^{2}}\lambda ^{2}+\frac{%
c^{2}(c^{2}M^{4}-6M^{2}-2{a}^{2} \epsilon )}{24a^{4}}\lambda
^{4}+\cdots \right]}_{F_{1}(\lambda)} + \\
&F_{II}\underbrace{\left[ \lambda +\frac{2-M^{2}c^{2}}{6a^{2}}\lambda ^{3}+\frac{%
24+c^{4}M^{4}-14M^{2}c^{2}-6{a}^{2} \epsilon c^{2}}{120a^{4}}\lambda
^{5}\cdots \right]}_{F_{2}(\lambda)}
\end{split}
\label{Gmu1}
\end{equation}%
where $F_{I}$ and $F_{II}$ are constants. We verify for different values of
constants, the series (\ref{Gmu1}) has an appropriate radius of
convergence. As an example, for $ \epsilon =1,a=2,n=1,M=1$ and $c=1$, the
series is convergent for $\left\vert \lambda \right\vert <2$ (Figure \ref{hofxii51}). The
recursion relation that we have used to derive (\ref{Gmu1}), is 
\begin{equation}
4k(k-1)%
\mathbb{Q}
_{k}-(k^{2}-3k+1)%
\mathbb{Q}
_{k-2}+%
\mathbb{Q}
_{k-4}=0 
\end{equation}
where $\mathbb{Q}_k$ is the coefficient of $\lambda^k$. 

\end{document}